\documentclass[journal]{IEEEtran}
\ifCLASSINFOpdf
\else
\fi

\usepackage{amsmath,graphicx,textcomp}
\usepackage{tabularx}
\usepackage{multirow}
\usepackage{url}
\begin{document}
\title{Deep CNN Framework for Audio Event Recognition using Weakly Labeled Web Data}

\author{Anurag Kumar,~\IEEEmembership{Student Member,~IEEE,}
        and Bhiksha Raj,~\IEEEmembership{Fellow, IEEE}
        % <-this % stops a space
%\thanks{M. Shell was with the Department
%of Electrical and Computer Engineering, Georgia Institute of Technology, Atlanta,
}

\maketitle

\begin{abstract}
The development of audio event recognition systems require labeled training data, which are generally hard to obtain. One promising source of recordings of audio events is the large amount of multimedia data on the web. In particular, if the audio content analysis must itself be performed on web audio, it is important to train the recognizers themselves from such data. Training from these web data, however, poses several challenges, the most important being the availability of labels : labels, if any, that may be obtained for the data are generally {\em weak}, and not of the kind conventionally required for training detectors or classifiers. We propose that learning algorithms that can exploit weak labels offer an effective method to learn from web data. We then propose a robust and efficient deep convolutional neural network (CNN) based framework to learn audio event recognizers from weakly labeled data. The proposed method can train from and analyze recordings of variable length in an efficient manner and outperforms a network trained with {\em strongly labeled} web data by a considerable margin. Moreover, even though we learn from weakly labeled data, where event time stamps within the recording are not available during training, our proposed framework is able to localize events during the inference stage. 
\end{abstract}

\begin{IEEEkeywords}
Audio Events, Web Audio Data, CNN
\end{IEEEkeywords}

% \fi
\IEEEpeerreviewmaketitle

\vspace{-0.2in}
\section{Introduction}
\IEEEPARstart{I}{n} \footnote{This is an updated version of the paper with results previously showcased in this thesis \cite{kumar2018acoustic} } last few years audio event detection (AED) has become an important research problem in the broad area of machine learning and signal processing. Audio event detection deals with the problem of recognizing and detecting sound events in audio or video recordings. One of the earliest application of AED was in surveillance \cite{atrey2006audio}, where the sound events such as \emph{gunshots, screaming} were of particular interest. Lately, its importance has been realized in a wide variety of areas such as content based retrieval of multimedia \cite{tong2014lamp,yu2014informedia}, human computer interaction \cite{ren2017sound}, wildlife monitoring \cite{stowell2016bird} to name a few. Moreover, as voice based assistants become more and more popular, the role of sounds to understand context and improve the responses will become critically important \footnote{http://blog-idcuk.com/sound-recognition-as-a-key-strategic-technology-for-artificial-intelligence/}. This is natural, as humans more often than not, rely on sounds to understand the environment and context around them. 

The early works used a variety of features and classifiers. GMM-HMM architectures \cite{zhuang2010real}, bag of audio words \cite{pancoast2012bag}, supervectors \cite{kumar2013event}, i-vector \cite{huang2013blind} features built over standard acoustic features such as Mel-Frequency Cepstral Coefficients (MFCCs). The success of deep learning in other audio tasks such as speech recognition \cite{hinton2012deep}, speech enhancement \cite{xu2015regression} led to the exploration of deep learning methods for sound event recognition as well. Convolutional Neural Networks (CNNs) have shown a lot of promise in sound event recognition \cite{piczak2015environmental, takahashi2016deep, hershey2017cnn}. However, a major constraint in large scale deep learning for AED is the availability of large scale dataset.  

Web has been a major source of large scale datasets for several tasks in the field of machine learning \cite{deng2009imagenet, thomee2016yfcc100m, gabrilovich2013facc1}. Multimedia data on the web are an important source of audio events as well. However, the recordings are usually unlabeled; to use them to train models for audio events one must label them with the time stamps that mark the temporal boundaries of the occurrences of different audio events. This form of labels, often referred to as ``strong'' labels, is a major bottleneck in large scale training of models for AED. Creating such labeled \emph{audio} data is very difficult as marking temporal boundaries of audio events is a time-consuming and expensive process. Often one must go back and forth several times in a recording to mark the beginnings and ends of audio events. 

\begin{figure}[t]
   \centering
   \includegraphics[width=0.5\textwidth]{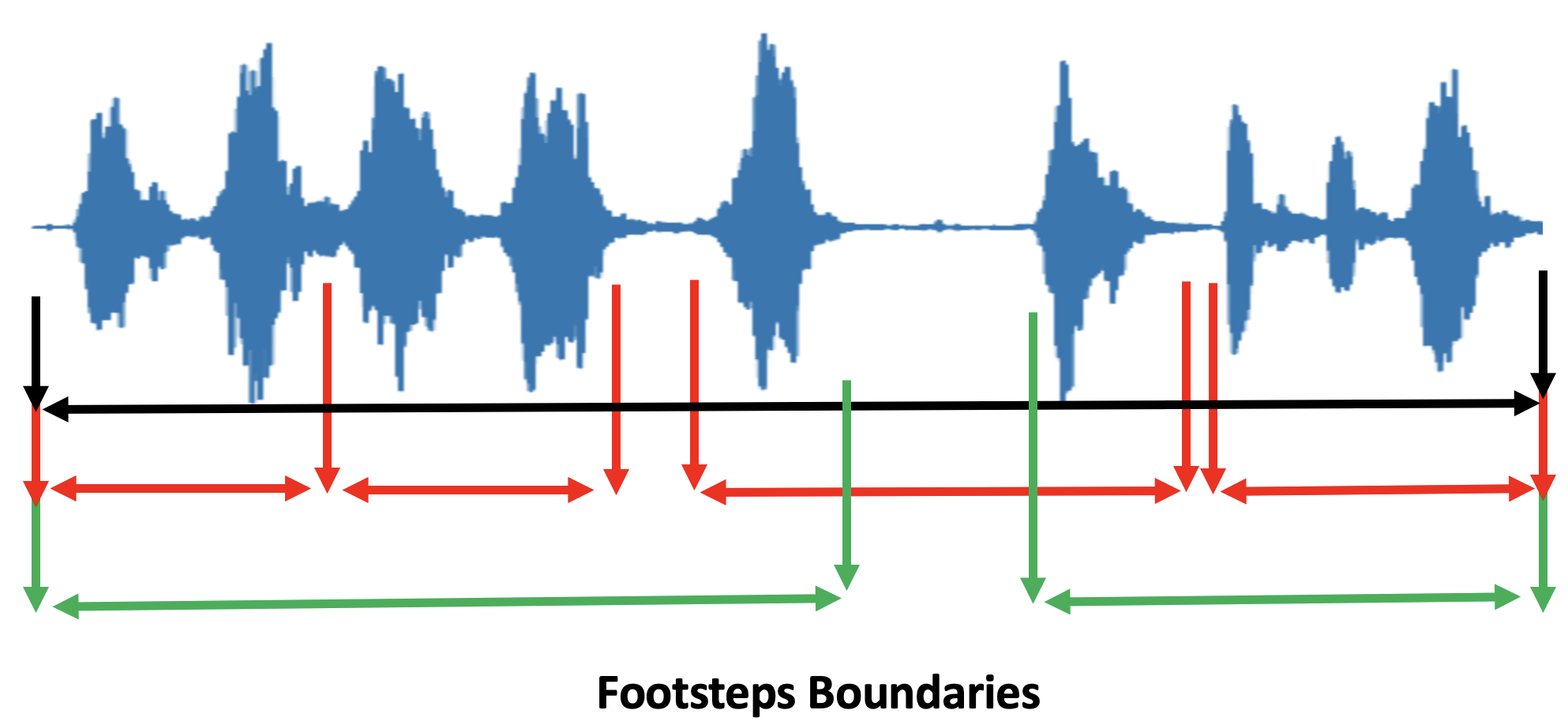}
     \caption{A recording of \emph{Footsteps}. Different annotators might mark different beginning and ends of each footstep in a given recording. Three such annotation example shown in black, red, and green colors.}
     \label{fig:footsteps}
\vspace{-0.20in}
\end{figure}

Moreover, for many sound events there is also a problem of {\em interpretation}. Consider for example, an audio recording of \emph{Footsteps} sounds shown in Figure \ref{fig:footsteps}. How many beginnings and ends should be marked for this recording: \emph{one}, \emph{two}, \emph{four}, or \emph{more}? Different annotators interpret the event differently and assign different beginning and ends to the footsteps.   Besides making the annotation task harder, this also creates unnecessary variability in the event exemplars, which is likely to confuse any algorithm that attempts to learn the underlying structure and composition of the audio event. 

To address these problems, \cite{kumar2016audio, anuragweakly} proposed that audio-event detects may be trained using {\em weakly} labeled data. In weakly labeled data only the presence or absence of the events is indicated; their actual location within the recordings is not marked. Since the time stamps need not be marked, it is much easier to label a recording for audio events. Moreover, the event when present in the recording is available to the learner in its entire natural form, without any assumption or bias about the beginning and ends which might get introduced by the annotator. Hence, the learning algorithm interprets the data in a more natural way. 

In this work, we argue that weak-label based learning approaches for AED gives us ways to exploit the vast and growing  amount of multimedia data on the web; which as we stated before are a rich source of audio events. Video-sharing websites such as \emph{YouTube} allows one to easily collect large amount of weakly labeled data for any given audio event. Hence, using web data the AED framework can be developed without requiring explicit labeling effort. We call such data as \emph{webly}  labeled data. This forms the primary motivation of this work where we aim to use web data for training audio event recognition models. Moreover, in order to be able to successfully deploy AED models for content-based indexing and retrieval of multimedia, and other real world applications, it is perhaps {\em essential} to train models on these data. 

Learning from webly labeled data presents several challenges. Web multimedia data are primarily consumer generated and are generally noisy. Recordings have inconsistent quality. The weak labels may encompass long audio segments, whereas the audio event of interest may itself be very short. Often, several sound events overlap, unlike the exemplars found in several of the audio event datasets such as \cite{salamon2014dataset,piczak2015esc}. A manually labeled large scale dataset for sound events is \emph{Audioset} \cite{gemmeke2017audio}. Labeling web data manually even for just weak labels can be very hard. Hence, it remains to be seen how multimedia data from web can be directly used for training audio event detection system without any manual labeling effort.

Some works have tried to develop CNN based approaches for audio-event recognition models using weakly labeled data \cite{hershey2017cnn, su2017weakly, xu2017attention}. Hershey {\em et al.} \cite{hershey2017cnn} explored using \emph{webly} labeled videos to experiment with well known CNN architectures for sound event recognition. They used audio from Youtube videos, which were automatically labeled by knowledge graphs for training and testing the architectures. 
Although they acknowledge the web data as being weakly labeled, the CNN models are trained in a fully supervised manner under the assumption that the labels are, in fact, strong. This, however, is not an efficient approach as it can result in a significant amount of label noise. An audio event, say \emph{door bell ringing}, may be present for only a few seconds in a recording which may be several minutes long, a fact that is ignored in assuming that the label is strong. Moreover, segment wise training of CNN is cumbersome and computationally inefficient. 

In this paper, we first propose a CNN-based framework which treats weak labels as \emph{weak} and does not make strong label assumptions during training. Our experimental results show that this is superior to training done under the strong-label assumptions. Moreover, our proposed framework can process input recordings of variable length, which makes the training and test process more efficient and convenient. The network design does not require a fixed size input and hence segmentation as a preprocessing step is not needed. The outputs at the segment level are directly produced by the network in one forward pass and the network design controls the segment and hop size.  The network can hence directly output temporal locations of events in the recording as well. Moreover, it is also computationally efficient as common operations are not repeated. We show the performance of the proposed method on first a manually annotated web data and then on \emph{webly} labeled data without any manual labeling.  
 
The rest of the paper is organized as follows. In Section \ref{sec:cnn} we describe our proposed approach for CNN based AED relying on weakly labeled data. In Section \ref{sec:expts} we describe the experiments and results. Lastly, we conclude in Section \ref{sec:conclusions}.  

\begin{figure*}[t!]
   \centering
   \includegraphics[width=0.85\linewidth]{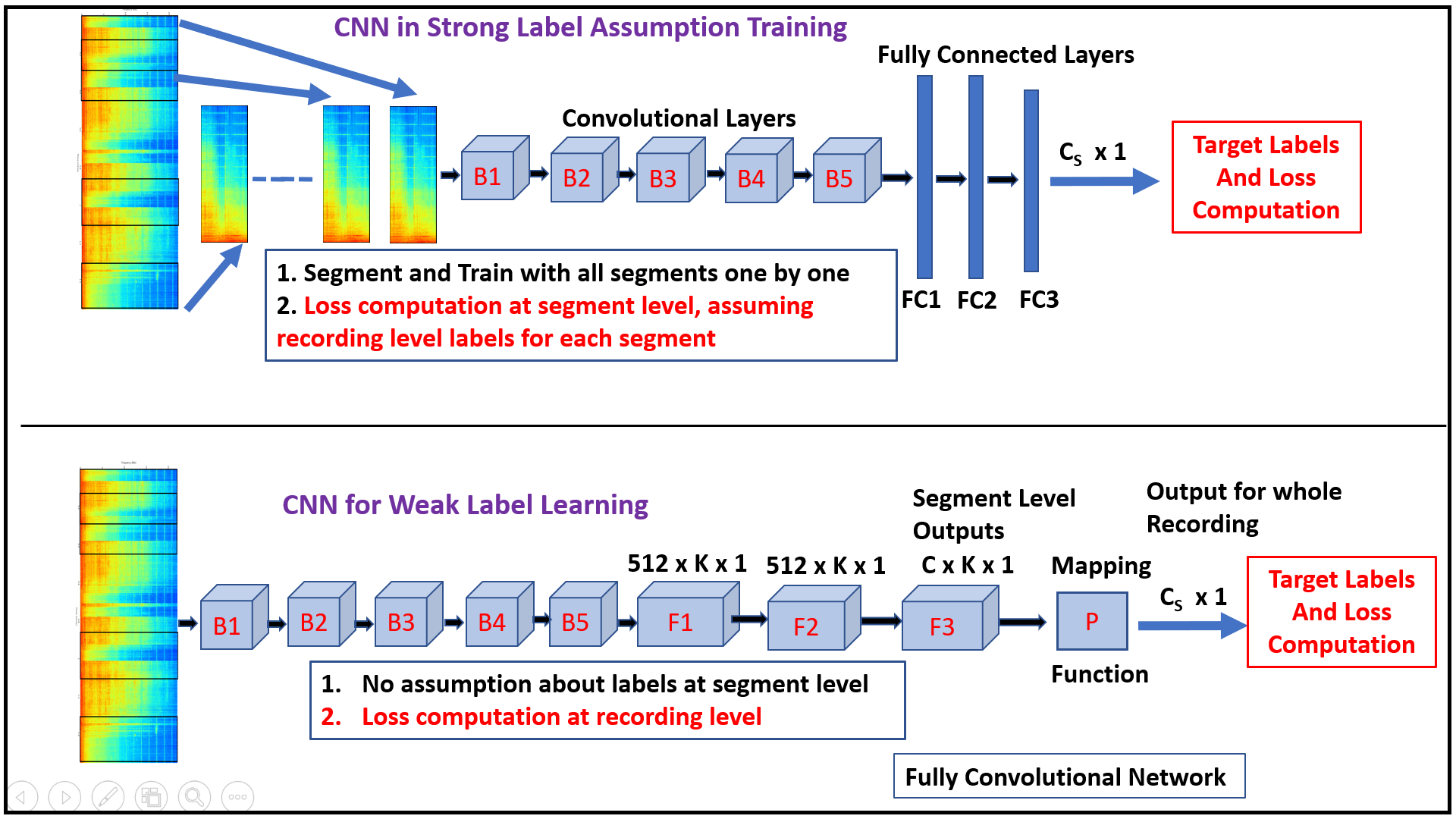}
   \vspace{-0.10in}
     \caption{\textbf{\emph{Top:}} Strong Label Assumption Training (SLAT, $\mathcal{N}_W$) - Trains CNN by assuming all segments of the recording contains the event, that is same label as full recording. \textbf{\emph{ Bottom:}} WALL-Net ($\mathcal{N}_{slat}$), weak label training of CNN. The input is full recording and network first produces outputs at recording level which is then mapped by the mapping function to produce recording level outputs. }
     \label{fig:framework}
\vspace{-0.2in}
\end{figure*}
\vspace{-0.05in}
\section{CNN FOR AUDIO EVENTS}
\label{sec:cnn}
Before going into the details of CNN models for AER, we would like to mention that we use logmel spectrograms as acoustic features to represent audio recordings. $128$ mel-bands are used.  From here on, a frame will refer to the single $128$ dimensional logmel vector. By ``recording" we imply the whole recording, represented by $X_R \in R^{N_R \times 128}$, the logmel representation of the full audio recording. $N_R$ is the number of frames in logmel representation of the audio recording. ``Segments" corresponds to small duration portions (say 1 or $1.5$ seconds) of the full audio recording. In the logmel space, segment will be $X_S \in R^{N_S \times 128}$, where $N_S$ is the number of frames corresponding to the segment duration. 

\vspace{-0.1in}
\subsection{CNN for Weakly Labeled Data}
CNN-based neural network architectures have been adopted for audio event recognition when strongly-labeled data are available \cite{piczak2015environmental, salamon2017deep,phan2016robust}. CNNs designed for strongly labeled data, consists of a few convolutional layers followed by fully connected layers. Since the data is strongly labeled, proper labels are available for each input and hence the loss function can be computed directly and network can be trained. The inputs are of fixed size (often obtained by segmenting into fixed duration segments) and labels for all inputs are known. 

For weakly labeled audio recordings on the other hand, we are given label for the whole recording. Consider a recording $\mathcal{R}$, with label vector $\vec{l}$. The label vector at index $i$, is $1$ if $C_i$ class is present, otherwise $0$. Since the time stamps for events are not known we cannot extract out event specific portion from $\mathcal{R}$ to train the network in supervised fashion. One simple way to handle this problem is through strong label assumption training. 

Let $S^{\mathcal{R}}_i, \,\, i = 1\,\,to\,\,K$ be small segments (say 1 second) obtained by chunking $\mathcal{R}$ into overlapping segments. $f$ be the network function such that $f(X)$ represents the network output for an input $X$ and $\mathcal{L}$ be the network loss function.  Let us assume that we have designed a network, which takes as input fixed duration segments $S^{\mathcal{R}}_i$, and produces output for each segment. Then the loss for the recording $\mathcal{R}$ can be computed as

\begin{equation}
\label{eq:strong}
Loss(\mathcal{R}) = \sum_{i=1}^{K} \mathcal{L}(f(S^{\mathcal{R}}_i),\vec{l})
\end{equation} 
In the above formulation, the network is trained assuming label $\vec{l}$ for all segments of $\mathcal{R}$. This naive approach, referred to as \emph{strongl label assumption training} (SLAT) was applied in \cite{hershey2017cnn}. This is the simplest approach to handle weakly labeled audio data. 

The MIL approaches for weakly label audio \cite{kumar2016audio}, however, does not make such assumption. In this paper, we propose to do the same. To achieve the same we propose the following loss 
\begin{equation}
\label{eq:weak}
Loss(\mathcal{R}) = \mathcal{L}(g(f(S^{\mathcal{R}}_1),\,\,f(S^{\mathcal{R}}_2),\,\, ...., \,\,f(S^{\mathcal{R}}_K))),\vec{l})
\end{equation} 

In Eq \ref{eq:weak}, the function $g()$, maps the segment level predictions ($f(S^{\mathcal{R}}_i)$) to recording level prediction. The loss is then computed using this recording level prediction with respect to the recording level label, $\vec{l}$. The mapping function $g()$ can in principle be any function which looks at predictions on smaller segments and then uses that knowledge to produce output for the whole recording. 

The idea follows the simple intuition that \emph{weak labels}, that is labels for the whole recordings, essentially  comes from presence or absence of events at lower levels, in this case we consider at the segment level. The role of the mapping function is to look at these segment level outputs and obtain the recording level output using them. For example, $g$ can be $max()$ or $avg()$ functions. The $max()$ function takes a ``max" over all segments, i.e the maximal output for a class over all segments is considered as output. The $avg()$ function takes the mean of the outputs across all segments as outputs for all classes. $g$ can also be a function with learnable weight parameters which does weighted combination of segment level outputs to produce recording level outputs. It can also be a recurrent neural architecture where the input sequence is the segment level outputs.  

The goal now is to design a network which takes as input the whole audio recording ($\mathcal{R} \in {N_R \times 128}$) and then first produce the output at the segment level and then map the segment level outputs to recording level. The network obviously should be able to handle input recordings of variable length. The next section describes the network architectures. 

\subsection{Network Architecture} 
The schema for the strong label assumption training and the proposed weak label training is shown in Figure \ref{fig:framework}. In SLAT, training is done by assuming that the labels for the segments are same as labels for the recording, leading to the loss function in Eq \ref{eq:strong}. Recordings are segmented and the passed to the network one by one, and labels for each segment is assumed to be same as label for the whole recording. 

The lower part of Fig. \ref{fig:framework} shows the schema for the proposed weak label learning. The network first produces output at segment level, $K$ is the number of segments for a certain input recording. $K$ depends on the size of input recording and the segment and hop size for which the network is designed. The mapping function $g$ then produces the recording level outputs using these segment level outputs. 

The layer blocks, B1 to B4 consist of two convolutional layers followed by a max pooling layer. The convolutional layers consist of filters of size $3 \times 3$, operated with a stride of 1. The input at all layers are padded, with padding size set to 1. The number of channels used in each layer within each block is as follows: \{B1: 32, B2: 64, B3: 128, B4: 256\}. The max pooling is done over a window of size $2 \times 2$ moving with a stride of 2. 

B5 block is similar to previous blocks except that it contains only 1 convolutional layer, with $256$ channel outputs, which is followed by the max pooling layer. Layers F1, F2 and F3 are convolutional layers with 512, 512 and $C$ number of channels respectively. F1 has convolutional filters of size $4 \times 4$; F2 and F3 has filters of size $1 \times 1$. Stride of $1$ is used without any padding on the inputs of these layers. ReLU activation is used in all convolutional layers except for F3. F3 is the segment level output layer with sigmoidal activation output. The segment level output layer is followed by the mapping function $g$. In this work, we use is either $max()$ or $avg()$ as the mapping function $g$.   

We will call our network \emph{WALL-Net} (Weak Audio Label Learning Network), $\mathcal{N}_{W}$. The network is fully convolutional and hence can handle recordings of variable length. The value of $K$ or number of segments clearly depends clearly on the size of input recording. 

Consider an input recording $\mathcal{R}$ represented by its logmel features and contains $1024$ logmel frames that is $\mathcal{R} \in R^{1024 \times 128}$. It can be easily computed that $K$ for this input will come out to be $29$,  for this recording will be $29$. The network is designed to produce outputs at every segment of size $128$ frames moving by $32$ frames. 

The segment layer outputs can be used for temporal localization of events. Since we can map the receptive field of the segment level output in the input and know the exact frames represented by each segment, we can obtain frame level outputs for each event. We illustrate this in the experiments and results section. Note that the fully covolutional nature of \emph{WALL-Net} allows us to handle audio recordings of variable length with ease. 

\vspace{-0.1in}
\subsection{Loss Function} 
\vspace{-0.05in}
Multi-label Problem: Often several audio events are simultaneously present in an audio recording. Hence, we design our network for multi-label training and prediction. The sigmoid output in the last layer can be considered as class specific posterior for any given input. Binary cross entropy function as shown in Eq \ref{eq:bce} is then used to compute loss with respect to each class. 
\begin{equation}
\label{eq:bce}
l(y_c,p_c) = -y_c*log(p_c) - (1-y_c)*log(1-p_c)
\end{equation}
In Eq \ref{eq:bce} $y_c$ and $p_c = \mathcal{N_S(X)}$ are target and network output for $c^{th}$ class respectively. The overall loss function is the mean of losses over all classes, Eq \ref{eq:lossfn} 
\begin{equation}
\label{eq:lossfn}
L(\mathcal{X},y) = \frac{1}{C} \sum_{c=1}^C l(y_c,p_c)
\end{equation}

\vspace{-0.1in}
\subsection{Multi-Scale Acoustic Features}
\vspace{-0.05in}
As stated before, we employ logmel spectra as acoustic features for training the network. For training the network, we extract logmel features at different FFT scales. The sampling rate for all recordings is 44100 Hz. Window (FFT) sizes used are 23ms (1024), 46ms (2048), and 92ms (4096). The hop size is fixed to 11.5ms (512) for all four window sizes and 128 mel-bands are used to extract mel spectra. 

During training the network is simply trained on all feature variants. Essentially,  the multiscale feature training serves as a data augmentation method. It exposes the network to different frequency representation which leads to better learning. During prediction mel spectra for the test recording at each FFT size is forwarded through the network and then the average output score for each class across all FFT sizes is considered as the final output. 

\vspace{-0.1in}
\section{Experiments and Results}
\label{sec:expts}
\vspace{-0.1in}
\subsection{Datasets}
\label{ssec:ds}
\vspace{-0.05in}
We consider two different web data sources in our experiments.

\noindent \textbf{Urbansounds (US)}:  Urbansounds \cite{salamon2014dataset} dataset gives us human annotated weakly labeled data. The source of audio recordings in this dataset is Freesound website \cite{freesound}. $10$ events, namely \emph{Air Conditioner, Car Horn, Children Playing, Dog Barking, Drilling, Engine Idling, Gunshot, Jackhammer, Siren and Street Music } were manually marked in the recordings. Partial time stamp information for each recording is provided in the dataset. We do not use that information and all experiments rely only on weak labels. A total of 1302 recordings, amounting to about $27$ hours is present in the dataset, with recording length varying from a few seconds to  around 10 minutes. The dataset comes pre-divided into 10 folds. We use the first 4 for training (533 recordings), next 2 (262 recordings) for validation and last 4 (502 recordings) as the testing set. 

\textbf{Youtube Training Set}:  The importance of weakly labeled learning lies in being able to obtain labeled data directly from web without any human labeling effort and be able to train robust AED models from these data. To show this, we collect training data for the 10 sound events directly from Youtube. 

Collecting videos from YouTube and automatically getting their weak labels is a challenge in itself. We propose a simple yet effective strategy to collect weakly labeled data from YouTube. We form the text query for searching on YouTube by adding the keyword ``sound" to the event name, e.g \emph{children playing sound}. This leads to a considerable improvement in the retrieved queries for any given sound event. We then select the top 125 videos under 10 minutes duration retrieved by YouTube and mark them to contain that event. The duration of the recordings vary from 0.6 seconds to \texttildelow 10 minutes, with an average duration of 2.1 minutes. The total audio data collected is around 48 hours. 

\textbf{Audioset Test Set}: US dataset is  a relatively clean dataset. Often it is required to recognize events in low quality consumer generated data, which are more like the noisy web data we use for training here. To test our approach on data of such form and nature, we used the ``Eval" set from Audioset dataset \cite{gemmeke2017audio}. The source of Audioset is also YouTube. $9$ events out of the 10 events considered, are present in Audioset (except \emph{street music}) and we will present results only for 9 events. A total of 761 test recordings exist with durations of 10 seconds for most cases. 
%We ensured that these test recordings are not part of our Youtube training set.  

Note that, due to mel spectra feature extraction at 3 different FFT scales, the total experimental data is 3 times in all cases, that is up to 144 hours of audio data in case of the YouTube training set.  

\subsection{Baseline and Experimental Setup}
\vspace{-0.05in}
We compare our proposed framework against SLAT \cite{hershey2017cnn}. $\mathcal{N}_{slat}$ is trained on recording segments where each segment is given the recording level label. The fully connected layers FC1, FC2 and FC3 has 512, 512 and $C$ number of nodes respectively. Hence, the network configuration matches our WALL-Net to make the comparison fair. 
During prediction the recording level score for each class is obtained by applying $max()$ or $avg()$ over segment scores. Finally, averaging across all 3 mel spectra features similar to the weak label case is done. 

Similar to \cite{hershey2017cnn}, we will use Area under ROC curves (AUC) and Average Precision (AP) as performance metrics \cite{fawcett2004roc, buckley2004retrieval}. We will report Mean AUC (MAUC) and Mean AP (MAP) over all classes as the overall metrics for comparison.

We use Pytorch\footnote{https://pytorch.org/} for implementation of our neural networks. The network is trained with Adam optimizer \cite{kingma2014adam}. Validation set is used for the learning rate and model selection across different epochs. 
\vspace{-0.15in}
\subsection{Urbansounds Results}
\begin{table}[t!]
\caption{Performance of $\mathcal{N}_W$ and $\mathcal{N}_{slat}$ and effect of multi-scale training}
\vspace{-0.1in}
\centering
\resizebox*{1.0\columnwidth}{!}{
\begin{tabular}{|c|c|c|c|c|}
\hline  
FFT scales used &\multicolumn{2}{c|}{MAP} & \multicolumn{2}{c|}{MAUC}\\ 
\cline{2-5}
in training &$\mathcal{N}_{slat}$&$\mathcal{N}_W$&$\mathcal{N}_{slat}$&$\mathcal{N}_W$\\
\hline
1024 & 0.641 & 0.715 & 0.899 & 0.930 \\
\hline
1024, 2048 & 0.662 & 0.734 & 0.905 & 0.927 \\
\hline
1024, 2048, 4096 & 0.697 & 0.750 & 0.905 & 0.935 \\
\hline
\end{tabular}
}
\label{tab:USmain}
\vspace{-0.10in}
\end{table}
Table \ref{tab:USmain} compares $\mathcal{N}_W$ and $\mathcal{N}_{slat}$. Table \ref{tab:USmain} shows that the proposed weak label learning network outperforms the strong label assumption training by a considerable margin. For networks trained on just 1024 point FFT scale features, $\mathcal{N}_W$ outperforms $\mathcal{N}_{slat}$ by $11.5\%$ in terms of MAP and $3.45\%$ in terms of MAUC. Augmenting the data by adding features at different scales is helpful for both $\mathcal{N}_W$ and $\mathcal{N}_{slat}$. For $\mathcal{N}_W$, the MAP goes up by $2.65\%$ by adding 2048 point FFT features and by $5.6\%$ when trained on features at all 3 scales. 

\begin{table}[t!]
\caption{Performance of $\mathcal{N}_W$ for different mapping functions}
\vspace{-0.1in}
\centering
\resizebox*{1.0\columnwidth}{!}{
\begin{tabular}{|c|c|c|c|}
\hline  
\multicolumn{2}{|c|}{MAP} & \multicolumn{2}{c|}{MAUC}\\ 
\hline
$g=max()$ & $g=avg()$ & $g=max()$ & $g=avg()$\\
\hline
0.700 & 0.750 & 0.904 & 0.935 \\
\hline
\end{tabular}
}
\label{tab:USmapping}
\vspace{-0.10in}
\end{table}

The results presented in Table \ref{tab:USmain} for $\mathcal{N}_W$ uses $avg()$ as mapping function $g$. Table \ref{tab:USmapping} compares the results for $\mathcal{N}_W$ using $max()$ and $avg()$ as mapping function. All 3 scales of features are used. We observe that $avg()$ mapping outperforms the $max()$ mapping. The $max()$ can be thought of as picking one segment for each class which contributes to the loss function computation and updates. The $avg()$ function on the other hand, allows all segments to contribute to the recording level outputs and hence loss and updates. Both mapping functions essentially can be thought of as having a weight vector, which is used to take the weighted combination of the segment outputs, to produce the recording level outputs. In case of $avg()$ this is $1/K$, $K$ the number of segments. In case of $max()$ it is a one hot vector, the index of $1$ corresponds to the segment which produces max output for a given class. One might expect that having a learnable weight parameter which learns the weights for combining the segment level outputs to obtain recording level outputs. In future we would try to explore this aspect of $\mathcal{N}_W$. 

\begin{table}[t!]
\centering
\caption{AP and AUC for each event using $\mathcal{N}_{slat}$ and $\mathcal{N}_W$  }
\resizebox*{1.0\columnwidth}{!}{
\begin{tabular}{|c|c|c|c|c|}
\hline  
Event &\multicolumn{2}{c|}{AP} & \multicolumn{2}{c|}{AUC}\\ 
\cline{2-5}
Name & $\mathcal{N}_{slat}$ & $\mathcal{N}_W$ & $\mathcal{N}_{slat}$ & $\mathcal{N}_W$\\
\hline
Air Conditioner & 0.507 & 0.477 & 0.807 & 0.817 \\
\hline
Car Horn & 0.693 & 0.834 & 0.884 & 0.957 \\
\hline
Children Playing & 0.774 & 0.879 & 0.951 & 0.978 \\
\hline
Dog Bark & 0.859 & 0.918 & 0.911 & 0.944 \\
\hline
Drilling & 0.669 & 0.622 & 0.931 & 0.922 \\
\hline
Engine Idling & 0.444 & 0.540 & 0.795 & 0.871 \\
\hline
Gunshot & 0.832 & 0.929 & 0.983 & 0.990 \\
\hline
Jackhammer & 0.685 & 0.703 & 0.940 & 0.939 \\
\hline
Siren & 0.703 & 0.694 & 0.902 & 0.954\\
\hline
Street Music & 0.800 & 0.907 & 0.949 & 0.978\\
\hline
Mean & 0.697 & 0.750 & 0.905 & 0.935 \\
\hline
\end{tabular}
}
\vspace{-0.15in}
\label{tab:Usclasswise}
\end{table}

Table \ref{tab:Usclasswise} shows event wise results for $\mathcal{N}_W$ and $\mathcal{N}_{slat}$. For most events $\mathcal{N}_W$ outperforms $\mathcal{N}_{slat}$. The difference is considerable for several events, upto $20 - 21\%$ for events such as \emph{Car Horn} and \emph{Engine Idling}. Interestingly, there are a couple of events, \emph{Air Conditioner} and \emph{Engine Idling} for which there is a small drop in performance and the network with strong label assumption training does better. Overall, however, training the network with the proposed method outperforms $\mathcal{N}_{slat}$ by around $7.6\%$. 

The segment size at which the outputs are predicted is another factor which can play a role in the performance. In the previous tables, the network was designed for segment size $128$ frames $\approx \,\, 1.5\,\, seconds$. Table \ref{tab:USsegsize}, compares the results for $\mathcal{N}_W$ designed for segment size of $128$ frames and of $64$ frames $\approx \,\, 0.75\,\, seconds$. The network with segment size $64$ frames is designed by changing the filter size in layer F1 to $2 \times 4$. Overall, the 128 frames segment size  performs better than 64 frames segment size by about $5.3\%$ in terms of MAP. However, the role of segment size can be event specific. Due to space constraints, we do not show results of all events. However, it is worth noting that for event such as \emph{Air Conditioner}, which has long term, more or less stationary characteristics, longer segment size is much better. 1.5 s segment size is better than 0.75 s segment size by more than $35\%$. At the same time, for event such as \emph{Engine Idling} and \emph{Siren} which have shorter salient characteristics, 0.75 s segment size is better than 1.5 s segment size by around $18\%$ and $9\%$ respectively.

\begin{table}[t!]
\caption{Performance of $\mathcal{N}_W$ designed for different mapping functions}
\vspace{-0.1in}
\centering
\resizebox*{1.0\columnwidth}{!}{
\begin{tabular}{|c|c|c|c|}
\hline  
\multicolumn{2}{|c|}{MAP} & \multicolumn{2}{c|}{MAUC}\\ 
\hline
128 (1.5 sec) & 64 (0.75 sec) & 128 (1.5 sec) & 64 (0.75 sec)\\
\hline
0.750 & 0.712 & 0.935 & 0.922 \\
\hline
\end{tabular}
}
\label{tab:USsegsize}
\vspace{-0.10in}
\end{table}
\begin{figure}[t]
   \centering
   \includegraphics[width=0.7\columnwidth]{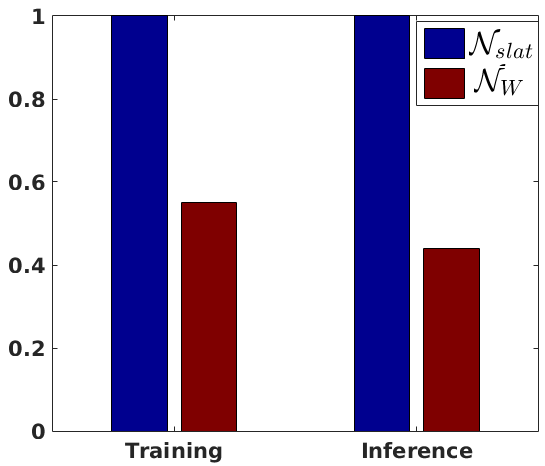}
     \caption{Comparison of Computational time for $\mathcal{N}_W$ and $\mathcal{N}_{slat}$. $\mathcal{N}_W$ is almost twice faster compared to $\mathcal{N}_{slat}$ during both training and inference.}
     \label{fig:timings}
\vspace{-0.20in}
\end{figure} 

\textbf{Computational Expense}: Figure \ref{fig:timings} shows the average training and inference time comparison for the two methods. $\mathcal{N}_W$ is over $45\%$ faster than $\mathcal{N}_{slat}$ during training and more than $55\%$ during inference. It shows that the proposed $\mathcal{N}_W$ is not only better and more convenient (no segmentation or other such preprocessing) but is also computationally much better. 

\vspace{-0.1in}
\subsection{YouTube Results}
\begin{table}[t!]
\centering
\caption{AP and AUC for each event using $\mathcal{N}_{slat}$ and $\mathcal{N}_W$  with YouTube Training}
\resizebox*{1.0\columnwidth}{!}{
\begin{tabular}{|c|c|c|c|c|}
\hline  
Event &\multicolumn{2}{c|}{AP} & \multicolumn{2}{c|}{AUC}\\ 
\cline{2-5}
Name & $\mathcal{N}_{slat}$ & $\mathcal{N}_W$ & $\mathcal{N}_{slat}$ & $\mathcal{N}_W$\\
\hline
Air Conditioner & 0.106 & 0.207 & 0.610 & 0.720 \\
\hline
Car Horn & 0.269 & 0.440 & 0.773 & 0.854 \\
\hline
Children Playing & 0.347 & 0.317 & 0.828 & 0.893 \\
\hline
Dog Bark & 0.704 & 0.800 & 0.944 & 0.977 \\
\hline
Drilling & 0.151 & 0.296 & 0.736 & 0.801 \\
\hline
Engine Idling & 0.371 & 0.424 & 0.773 & 0.796 \\
\hline
Gunshot & 0.589 & 0.755 & 0.795 & 0.927 \\
\hline
Jackhammer & 0.101 & 0.135 & 0.510 & 0.668 \\
\hline
Siren & 0.753 & 0.730 & 0.852 & 0.805\\
\hline
Mean & 0.377 & 0.456 & 0.757 & 0.827 \\
\hline
\end{tabular}
}
\vspace{-0.15in}
\label{tab:ytmain}
\end{table}

Table \ref{tab:ytmain} shows the performance for the networks when trained on webly labeled YouTube recordings \ref{ssec:ds}. The manually labeled Audioset data serves as test set. The results correspond to training and testing with all 3 feature scales. Once again we see that $\mathcal{N}_W$ outperforms $\mathcal{N}_{slat}$. For several events such as \emph{Air Conditioner}, \emph{Car Horn}, \emph{Drilling} improvements in the range of $63\%$ to $96\%$ in relative terms for AP can be seen. MAP over all events improves by around $21\%$. This gain is considerably higher than that observed in US dataset. This   shows that for webly labeled data weak label training is crucial and strong label assumptions should not be made.  

Note that, the overall performance numbers are much lower compared to results on the Urbansounds dataset, where the recordings are relatively cleaner and the presence of noise and other events are relatively low and label noise is absent as the labels are manually obtained. The \emph{YouTube} training and test set on the other hand, are much more noisy and contains other audio events which makes them more challenging as far as recognition is concerned. The training set most likely contains label noise as no manual checking was performed. 

We compare the performance of the model trained on Urbansounds dataset to that of the model trained on \emph{YouTube} on \emph{Audioset} test set. Table \ref{tab:ytus} shows this comparison. We notice that the \emph{YouTube} training set which is only \emph{webly} labeled, without any manual labeling effort outperforms the network trained on Urbansounds dataset. A relative improvement of around $9\%$ in terms of MAP and $2.5\%$ in terms of MAUC is observed. For several events the difference is much higher, for example for \emph{Car Horn} and \emph{Dog Bark} the improvement is by more than $57\%$ and $31\%$ respectively. However, there are events such as \emph{Children Playing} and \emph{Siren}, where the network trained on Urbansounds set is better compared to that trained on \emph{Audioset}. Overall, we can conclude that for recognizing events in unstructured recordings such as those on YouTube, which also resembles more to the kind of data we will see in real life situations, training on such data obtained from YouTube can be very useful.

\begin{table}[t!]
\centering
\caption{AP and AUC for each event using $\mathcal{N}_{W}$ trained on Urbansound and on \emph{YouTube} training set. }
\resizebox*{1.0\columnwidth}{!}{
\begin{tabular}{|c|c|c|c|c|}
\hline  
Event &\multicolumn{2}{c|}{AP} & \multicolumn{2}{c|}{AUC}\\ 
\cline{2-5}
Name & $\mathcal{N}_{W} $ (US) & $\mathcal{N}_W$ (YT) & $\mathcal{N}_{W}$ (US) & $\mathcal{N}_W$ (YT)\\
\hline
Air Conditioner & 0.155 & 0.207 & 0.713 & 0.720 \\
\hline
Car Horn & 0.280 & 0.440 & 0.816 & 0.854 \\
\hline
Children Playing & 0.631 & 0.317 & 0.937 & 0.893 \\
\hline
Dog Bark & 0.610 & 0.800 & 0.928 & 0.977 \\
\hline
Drilling & 0.215 & 0.296 & 0.777 & 0.801 \\
\hline
Engine Idling & 0.373 & 0.424 & 0.738 & 0.796 \\
\hline
Gunshot & 0.699 & 0.755 & 0.888 & 0.927 \\
\hline
Jackhammer & 0.113 & 0.135 & 0.586 & 0.668 \\
\hline
Siren & 0.804 & 0.730 & 0.858 & 0.805\\
\hline
Mean & 0.418 & 0.456 & 0.805 & 0.827 \\
\hline
\end{tabular}
}
\vspace{-0.15in}
\label{tab:ytus}
\end{table}

\subsection{Temporal Localization} 
Even though we learn from weakly labeled data, where temporal information is not available during training, it is often desirable to locate events during the inference stage. The previous results show recognition at the recording level. Computing objective measures for temporal localization requires ground truth locations of events in the recording. Neither Urbansounds test set nor Audioset test properly provides ground truth locations of events. Urbansounds provides only partial ground truth locations, again not allowing us to compute any objective measures.

\begin{figure}[t]
   \centering
   \includegraphics[width=0.9\columnwidth]{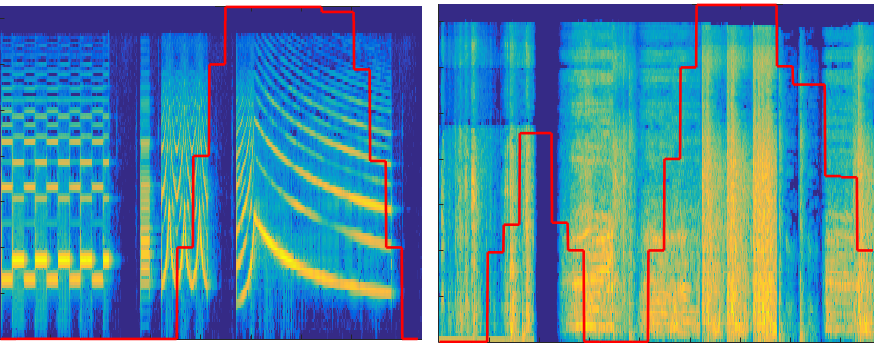}
     \caption{Temporal Localization Examples. Left: An Example of Siren Sound, Right: An Example of Gunshot Sound.}
     \label{fig:temporal}
\vspace{-0.20in}
\end{figure}

The segment level predictions at layer F3 isn $\mathcal{N}_W$ can be easily mapped back to the frame level prediction, since we can map back each segment level output to the receptive fields in input recordings. Figure \ref{fig:temporal} shows a couple of examples of localization as given by our method. The red line shows the output activation corresponding to the sound event as it changes with time. Note that as the event starts the activation goes up for the duration of the event.  

\vspace{-0.1in}
\section{Conclusions}
\label{sec:conclusions}
In this paper, we proposed a deep CNN  based framework to learn audio event recognition using web data.  We showed it is possible can collect weakly labeled data directly from web and then train audio event recognition models using that. Our proposed CNN based learning framework nicely incorporates  weakly labeled nature of the audio data. It outperforms training mechanism which makes strong label assumption by a considerable margin. Moreover, the proposed framework can efficiently handle recordings of variable length during training as well as testing. No pre-processing to segment the recording into fixed length is required. Besides a better and smoother implementation process this is also computationally more efficient. Our proposed framework can perform temporal localization as well using the segment level outputs. This can be useful for applications where we need to know where the event occurred in the recording. 
%Webly labeled data, however, comes with the problem of label noise. We plan to investigate problems associated with \emph{label noise} in webly labeled in future. 

% or


%\appendices

%\section{}

%\section*{Acknowledgment}

\ifCLASSOPTIONcaptionsoff
  \newpage
\fi

\bibliographystyle{IEEEtran}
\bibliography{references}

\end{document}